\documentclass{article}

\usepackage{graphicx}        
\usepackage{amssymb}        
\usepackage{color}           
\usepackage{geometry}
\usepackage{url}

\newcommand{\arcsec}{arcsec}
\newcommand{\farcs}{^{\prime\prime}}
\chardef\us=`\_
\newcommand{\citep}{\cite}

\title{The deformation of an erupting magnetic flux rope in a confined solar flare}
\author{Ruisheng Zheng$^{1, 2, 3, \star}$, Yihan Liu$^{1}$, Liang Zhang$^{1}$, Yang Liu$^{1}$, Changhui Rao$^{4, 5, 6}$, Qing Lin$^{7}$, \\Zhimao Du$^{7}$, Libo Zhong$^{4, 5}$, Huadong Chen$^{3}$, Yao Chen$^{1, 2}$}

\begin{document}

\maketitle

1 Institute of Space Sciences, Shandong University, Weihai 264209, China\\
2 Institute of Frontier and Interdisciplinary Science, Shandong University, Qingdao 266237, China\\
3 CAS Key Laboratory of Solar Activity, National Astronomical Observatories, Beijing 100012, China\\
4 Institute of Optics and Electronics, Chinese Academy of Sciences, Chengdu 610209, China\\
5 The Key Laboratory on Adaptive Optics, Chinese Academy of Sciences, Chengdu 610209, China\\
6 University of Chinese Academy of Sciences, Beijing 100049, China\\
7 Shanghai Astronomy Museum, Shanghai 201306, China

\begin{abstract}
Magnetic flux ropes (MFRs), sets of coherently twisted magnetic field lines, are believed as core structures of various solar eruptions. Their evolution plays an important role to understand the physical mechanisms of solar eruptions, and can shed light on adverse space weather near the Earth. However, the erupting MFRs are occasionally prevented by strong overlying magnetic fields, and the MFR evolution during the descending phase in the confined cases is lack of attention. Here, we present the deformation of an erupting MFR accompanied by a confined double-peaked solar flare. The first peak corresponded to the MFR eruption in a standard flare model, and the second peak was closely associated with the flashings of an underlying sheared arcade (SA), the reversal slipping motion of the L-shaped flare ribbon, the falling of the MFR, and the shifting of top of filament threads. All results suggest that the confined MFR eruption involved in two-step magnetic reconnection presenting two distinct episodes of energy release in the flare impulsive phase, and the latter magnetic reconnection between the confined MFR and the underlying SA caused the deformation of MFR.

\end{abstract}

\section{Introduction}
Magnetic flux ropes (MFRs) are regarded as coherent sets of twisted magnetic field lines that wrap around their central axis, and their existence is evidenced by filaments/prominences, coronal cavities, and sigmoid structures in different emission lines. It is generally believed that MFRs are core structures of various explosive solar activities. The evolution of MFRs plays an important role to understand the physical mechanisms of solar eruptions, and can be one main driver for adverse space weather near the Earth \citep{chen17, cheng17, gibson18, Wang19, liu20}. After the destabilization due to kink instability \citep{torok05}, torus instability \citep{Kliem06, aulanier10}, breakout reconnection \citep{antiochos99}, or tether-cutting reconnection \citep{Moore01}, a rising MFR will stretch the overlying magnetic field lines, and lead to magnetic reconnection in the current sheet below, and finally cause a solar flare that releases free magnetic energy \citep{masuda94, priest02}.

In the 3-dimensional standard flare model \citep{aula12, janv13}, the MFR is surrounded by a quasi-separatrix layers (QSLs) where the magnetic field connectivity changes significantly \citep{demoulin96}. When magnetic field lines pass through the QSLs, successive reconnection will occur and result in the obvious slipping motion along the QSLs \citep{demo06, aula06}. Due to slipping magnetic reconnection, the footpoints of QSLs can be indicated by the observed flare ribbons that always show a form of double J-shape structure. Hence, the eruption of the MFR surrounded by QSLs is always accompanied by the slipping motions of flare ribbons and the formation of hook structures at both ends of flare ribbons \citep{Li15, Zheng16}.

Besides of the successful eruptions, the rising MFRs can also be prevented to escape and cause confined eruptions \citep{ji03, Liu09, Guo10, Song14, Zheng17, Netzel12, Kushwaha15, Mitra22}. As disclosed by observations and simulations, the failure of MFR eruptions are predominantly due to the confinement of strong overlying magnetic fields as a cage \citep{Liu16, amari18, Yang18, Zheng19}. Consequently, for a confined eruption, the rising MFR experiences a deceleration and possibly falls back to the solar surface after its stop at a certain height \citep{ji03, Yang19}.

As that in successful eruption, the development of a confined MFR is part of the lifetime of the MFR, and is also important for understanding the physical mechanisms of solar eruptions. However, the developments of MFRs in confined eruptions are lack of attention. In this Letter, we focus on the evolution of an erupting MFR in a confined double-peaked flare, and find the deformation of the confined MFR by the external magnetic reconnection with underlying sheared arcades.

\section{Observations}
The confined eruption occurred in NOAA AR 12860 ($\sim$S29W23) on 2021 August 29, and was linked to a double-peaked flare of C7.4 class (SOL2021-08-29T00:44). The main observations of the eruption are from the Atmospheric Imaging Assembly (AIA; \cite{lemen12}) onboard the Solar Dynamics Observatory (SDO; \cite{pesnell12}), and the AIA images has a pixel size of 0$\farcs$6 and a cadence of 12 s. We also used observations from the Extreme Ultraviolet Imager (EUVI; \cite{Howard2008})-A on the the Solar Terrestrial Relations Observatory(STEREO; \cite{Kaiser2008}). The EUVI images have a pixel size of 1.58$\arcsec$, and their cadences are 2.5 minutes for the 195~{\AA}. The X-ray images from the X-Ray Telescope (XRT; \cite{Golub07}) aboard the Hinode \citep{Kosugi07} are collected to determine the X-ray structures in AR 12860, with a pixel size of $\sim$1$\arcsec$. The emission properties of the eruption is also investigated with the differential emission measure (DEM) method that employs the {\it xrt\_dem\_iterative2.pro} in SolarSoftWare package \citep{Cheng12, Song14}. In DEM method, the EM maps at different temperature ranges are obtained by a set of AIA images in six channels (i.e., 94, 131, 171, 193, 211, and 335~{\AA}).

The magnetic field evolution of AR 12860 is examined by full-disk magnetograms and intensity images from the Helioseismic and Magnetic Imager (HMI; \cite{Scherrer12}) also onboard SDO, with a cadence of 45 s and pixel size of 0$\farcs$6. The filaments in AR 12860 were well captured by the H$\alpha$ filtergrams from the Solar Magnetic Activity Research Telescope (SMART; \cite{UeNo04}) at Hida observatory, with a pixel size of $\sim$1$\arcsec$. Moreover, the details of the sunspots and filaments are complemented by the high-resolution images in H$\alpha$ and TiO from the Educational Adaptive-optics Solar Telescope (EAST; \cite{Rao22}) that was newly built in 2021 July at Shanghai Astronomy Museum, and their pixel resolutions are 0$\farcs$12 for H$\alpha$ and TiO.

To check the configuration of the confined eruption, the nonlinear force-free field (NLFFF) modeling \citep{wheat00, wiege04} is also utilized to construct the coronal structures by setting the HMI photospheric vector magnetogram as the bottom boundary. The NLFFF extrapolation is performed in a box of 116$\times$116$\times$117 uniformly distributed grid points with $\Delta x = \Delta y = \Delta z = 2 \arcsec$, and then we calculate the squashing factor $Q$ \citep{demoulin96, titov02} of the extrapolated magnetic field \citep{Guo17}. The radio dynamic spectra associated with the eruption are obtained from the network of Compound Astronomical Low frequency Low cost Instrument for Spectroscopy and Transportable Observatory (CALLISTO; {\url{https://www.e-callisto.org/}}).

\section{Results}
\subsection{Eruption and Deformation of MFR}
Figure 1 shows the overview of AR 12860 and its magnetic field evolution during the eruption. In the EAST TiO image (panel (a)), it is clear that AR 12860 is divided into two groups of sunspots, the preceding one (the orange box) and the following one (the pink box). The high-resolution image of EAST H$\alpha$ (panel (b)) clearly displays that a slender sigmoidal filament suspended over the AR (green arrows), and some shorter filament threads (the orange arrow) lay beneath the western portion of the sigmoidal filament. The HMI intensity images and magnetograms (panels (c)-(f)) confirm that the preceding group is dominated by negative polarities, and the following group is comprised of predominant positive polarities and parasitic negative polarities. Note that one parasitic sunspot (NS1, the blue arrow) emerged in the west of following group and interacted with surrounding positive polarities, which distorted the polarity inverse line in the following group. In addition, NS1 also moved westwards as well as another negative sunspot (NS2, the red arrow) in the preceding group. Along the shifting direction (the green line in panel (e)), the time-distance plot (panel (g)) clearly shows the sunspot movements with a speed of $\sim$100 m s$^{-1}$ for NS1-NS2. The magnetic flux variations (from 12:00 on August 28 to 06:00 UT on August 29) in the source region of the eruption (red boxes in panels (d) and (f))) is shown in panel (h). The unsigned negative flux kept increasing with a net increase of $\sim6 \times 10^{20} Mx$, and the positive flux continuously decreased with a net decrease of $\sim5 \times 10^{20} Mx$. It likely indicates an emergence and cancellation of magnetic flux that continued through the period of the eruption (from 00:20-00:50 UT, the green-shaded area in panel (h)). The magnetic flux emergence and cancellation is consistent with the intrusion of NS1 and the fading of the positive polarities in the source region (panels (d) and (f)).

During the continuous magnetic activities and sunspot displacement, an MFR formed in the AR and subsequently erupted (Figure 2 and Animation 1). The MFR exhibited a transient sigmoid in high-temperature channels, similar to the case in Kharayat et al. (2021), and its asymmetry was indicated by that the western portion was much brighter than the eastern portion (panels (a)-(b)). In the low corona and the chromosphere (panels (c)-(d)), the western portion of the MFR was replaced {\bf by} a twisted structure and a filament (green arrows), and the eastern portion was hardly distinguished. At $\sim$00:25 UT on August 29, the asymmetric MFR erupted and resulted in a C7.4 flare. Unfortunately, the eruption was seriously restricted by groups of overlying AR loops (panel (c)). The flare induced a J-shaped ribbon and an L-shaped ribbon (the curves in panel (g)). The L-shaped ribbon exhibited an obvious clockwise slipping motion (the curved dashed arrow), and a hook structure (H1) appeared in the extending end (the cyan arrow in the panel (h)). According to the 3-dimensional flare model \citep{aula12, janv13}, the J- and L-shaped ribbon represented footpoints of QSLs, and the appearance of the unusual L shape was possibly due to the asymmetric eruption. The slipping motion of the L-shaped ribbon resulted from magnetic reconnection successively occurring below the erupting MFR.

Interestingly, the L-shaped ribbon sequentially experienced a reversal slipping motion in a few minutes (Figure 3 and Animation 2). The anti-clockwise slippage led a new hook structure (H2; the cyan arrow) in the east of the L-shaped ribbon, accompanied by the darkening of the H1. Meanwhile, an underlying sheared arcade (SA) appeared in the western part of the AR (blue arrows). Intriguingly, during anti-clockwise slippage, some filament threads (the green arrow) over the SA slowly moved southwards, and the SA became much bright and showed some flashings (the pink arrows). Moreover, it is evident that some plasmoids were released from the SA in AIA 193~{\AA} (the red arrow in panel (f)), similar with hot blobs reported in \cite{Kushwaha15}. It likely indicates magnetic reconnections between the filament threads and SA. Along the selected path through two hook structures (S1, the green curve in panel (a)), the time-distance plots show the bi-directional slipping motions in AIA 304 and 171~{\AA} (panels (g)-(h)). The clockwise slippage was prominent with a speed of $\sim$80 km s$^{-1}$ (the green dotted lines). The reversed slippage was weaker and slower with a speed of $\sim$10 km s$^{-1}$ (the green dashed lines). Note that the reversed slippage closely followed with the movement of filament threads (the white dashed line in panel (g)), which indicates a close relationship between the reversed slippage and the movement of filament threads. Along the eruption direction (S2, the red line in panel (c)), the time-distance plot shows the evolution of the MFR (panel (i)). It is clear that the MFR first propagated a distance of $\sim$50 Mm at a speed of $\sim$150 km s$^{-1}$ (the white dotted line) and began to fall down at $\sim$00:30 UT with a speed of $\sim$30 km s$^{-1}$ (the white dashed line). The falling of the MFR is consistent with the shifting of filament threads in AIA 304~{\AA}.

\subsection{Two episodes of Magnetic Reconnection}
The eruption is also shown in the limb view from EUVI-A and in composite images in AIA 171 (red), 193 (green), and 94 (blue)~{\AA} (Figure 4 and Animation 3). In the limb view (panels (a)-(b)), the enveloping field lines in AR 12860 were nearly intact through the eruption, which possibly indicates a very limited expansion of the enveloping loops. In the composite images (panels (c)-(f), the high temperature component in 94 ~{\AA} (blue) shows clearly the evolution of the MFR from a transient sigmoid to a diffuse structure (red arrows) that was restricted in the limited space experiencing a much less expansion. Interestingly, the flashings (the blue arrow) on a sheared arcade (SA) was surrounded by the diffuse high-temperature structure of the MFR, which likely indicates the occurrence of the interaction between the expanding MFR and with the underlying SA in the limited cage. After the interaction, the newly-formed coronal loops (white arrows) became visible after the cooling of the diffuse high-temperature structure ($\sim$ 01:20 UT).

The evolution of the flare ribbons is shown in Figure 5. The J- and L-shaped flare ribbons were obvious in AIA 1600~{\AA} (panel (a)), and their contours are superposed on magnetic polarities of HMI magnetograms (panels (c) and (e)). For the eruption region (the dashed box in panel (a)), the intensity curves in AIA 1600, 304, and 94~{\AA} (panel (b)) clearly reveal two peaks at $\sim$00:30 and $\sim$00:50 UT. It is consistent with the double peaks (dotted lines) of the GOES 1-8~{\AA} soft X-ray (SXR) flux and its derivative curve that indicate the Neupert effect of the non-thermal and thermal flare emissions \citep{neupert68, Qiu21}. Hence, the eruption indeed involved two episodes of magnetic reconnection. The reconnected flux and the reconnection rate for the positive and negative magnetic flux, estimated by the amount of magnetic flux swept by the flare ribbons in the AIA 1600~{\AA}, are shown with blue and red curves in panels (d) and (f). The maximums of the reconnected flux through two episodes of reconnection are separately $\sim2 \times 10^{20} Mx$ and $\sim4 \times 10^{19} Mx$. Around the peak time of two episodes of magnetic reconnection, two Type III radio bursts (arrows in panel (g)) were recorded in the solar radio dynamic spectra from CALLISTO.

Figure 6 shows the eruption in EM maps in different temperature ranges. At two peak time of the double-peaked flare, the strongest emissions separately reveal the post-flare loops (yellow arrows) and the SA (pink arrows) with flashings. The post-flare loops are predominant in the high-temperature ranges (left panels). The SA with flashings are dominated in the low-temperature range (right panels). Note that the loops (white arrows in panels (f)) in 10-25 MK are consistent with the newly-formed loops in Figure 4(f), which possibly indicates that they were the productions of magnetic reconnection related to flashings. Hence, the eruption indeed involved two episodes of magnetic reconnection.

\section{Conclusions and Discussion}
Combining with high-quality observations from EAST, HMI, and AIA, we report a confined eruption of an asymmetry MFR. The eruption was closely related to the continuous magnetic field activities (magnetic flux emergence and cancellation) and the sunspot displacement (Figure 1). The confined eruption was also asymmetry, which led to an L-shaped flare ribbon with a hook in the predominant eruption direction and a normal J-shaped flare ribbon in the opposite direction (Figure 2). Intriguingly, the L-shaped flare ribbon showed a reversal slipping motion, and was accompanied by the slow shifting of filament threads and the flashings of an underlying SA, which likely indicates another magnetic reconnection (Figure 3). Two episodes of energy release process for the confined eruption were indicated by AIA composite images and EM maps, and were also evidenced by the double peaks in the EUV, SXR, reconnected flux curves, and two type III radio bursts (Figure 4-6).

For the confined eruptions, it is believed that the upward force for rising instable MFRs will finally be balanced by the strong confinement of overlying field lines \citep{ji03, torok05, amari18, Netzel12, Mitra22}. If an twisted MFR meets the overlying field lines, the external magnetic reconnection possibly take places, and thus the twist of MFR can be transferred to a larger coronal magnetic system \citep{Gary04, DeVore08}. On the other hand, the confined MFR will stop at a certain height or fall back to the solar surface, when there is no interaction with the overlying field lines. Moreover, if one filament exists inside, the dark filament material can be seen in the low-temperature wavelength during the falling of the MFR \citep{ji03}. Due to the strong confinement, the confined MFR fell down (Figure 3(i)) and possibly met with an underlying SA, indicated by the slow shifting of the low-temperature counterpart (filament threads), the flashings, and the plasmoids (Figure 3(d)-(f)). Moreover, along the L-shaped flare ribbon, the rising of the erupting MFR caused the clockwise slipping motion, while the reversal slipping motion occurred during the falling of the confined MFR (Figure 3(g)-(i)), which also infers the occurrence of magnetic reconnection. Hence, in this Letter, the external magnetic reconnection occurred between the confined MFR and the underlying SA.

Interestingly, the flare ribbons showed both the clockwise and anti-clockwise slippage. The speed of the clockwise slippage is $\sim$80 km s$^{-1}$, in the range from several tens to $\sim$100 km s$^{-1}$ for the typical slipping motion in the 3-dimensional flare model \citep{Li15, Zheng16}, whereas the speed of the anti-clockwise slippage is only $\sim$10 km s$^{-1}$ (Figure 3(g)-(f)). The significant difference on the slippage speed is likely resulted from two kinds of magnetic reconnection, consistent with the different reconnection flux and rates (Figure 5 (d) and (f)). The fast clockwise slippage is likely induced by the successive reconnection between two groups of arcades overlying the rising MFR in the impulsive eruption stage, and the slower anti-clockwise slippage possibly indicates an intermittent reconnection between the confined falling MFR and the underlying SA. Hence, it is possible that the reversal slippage represents the shrinking of the footprints of the confined MFR following the latter external reconnection.

The double-peaked feature of the flare is the indicator of two-step magnetic reconnection involved with the confined eruption. The first peak corresponded to the internal magnetic reconnection between the stretching anti-parallel field lines in a standard flare model, and the second peak was closely associated with the external magnetic reconnection between the confined MFR and the underlying SA. In the derivative of SXR flux (Figure 4(g)), the first peak is much stronger than the second peak, which possibly means that the non-thermal emissions in the internal magnetic reconnection were much intense than that in the external magnetic reconnection.

It is reasonable that two-step magnetic reconnection releases energy in two episodes in the form of double peaks of the SXR flare. However, it is not often that the eruption involved with two-step magnetic reconnection must be accompanied by a double-peaked SXR flare. In some cases with M- and X-class flares \citep{Hao12, Gou17, Zou19, Zheng21}, the SXR flux only exhibited a short bump or nothing in the impulsive rising phase that lasted for a few minutes. The impulsive phase of the double-peaked flare in this Letter had a period of $\sim$24 minutes in GOES SXR and $\sim$34 minutes in AIA 94~{\AA} (yellow and cyan shades in Figure 5(b)), which is similar to that in \cite{Dumbovic21}. Hence, we suggest that, for the flare involving with two-step magnetic reconnection, the longer the duration of the impulsive phase is, the clearer the profile of the double peaks becomes.

In addition, we check the configuration of the source region before the eruption (00:00 UT) with the NLFFF model in Figure 7. The extrapolated fields are superposed on HMI vector magnetogram in the side view (left) and on contours of $Q$ value at the photospheric surface in the top view (right). The orange group of filed lines represents the MFR, and two groups of overlying loops are indicated by the blue and red field lines. The high $Q$ value beneath the MFR center indicates the possible sites of the internal reconnection, though the value is much less than that in earlier reports \citep{Mitra20, Joshi21}. The forward and reverse slipping motions are also superimposed on the counters of $Q$ value (yellow and cyan dashed arrows).

Based on the above results and discussions, we propose a possible scenario for the confined eruption in the schematic drawings in Figure 8. At the eruption stage (panels (a)-(b)), the rising MFR (red lines) stretched two groups of overlying loops (blue and green lines), and initiated the internal magnetic reconnection (the star symbol) that occurred beneath the rising MFR between two groups of overlying loops, following the 3-dimensional flare model. As a result, magnetic reconnection induced the lower post-flare loops and the higher and longer loops wrapping the rising MFR (blue-green lines), and J-shaped and L-shaped flare ribbons (purple curves). Due to the asymmetric eruption, the L-shaped ribbon appeared and exhibited the slipping motion (the clockwise arrow). The eruption was confined by the higher overlying strong cage (orange). At the post-confinement stage (panels (c)-(d)), the SA (the black line) appeared beneath the confined MFR, and the external magnetic reconnection (the star symbol) occurred between the MFR and the underlying SA. The external reconnection was accompanied by the reversal slipping motion (the anti-clockwise arrow) of the L-shaped ribbon, and the MFR finally deformed as two newly-formed loops (pink-black lines).

Most of studies for MFRs focus on the evolution of the formation and eruption \citep{ji03, torok05, amari18, Netzel12, Mitra22}, and the MFR evolution during the descending phase in the confined case, as the primary aim of this Letter, is lack of attention. In summary, the confined eruption of the MFR involved with two-step magnetic reconnection relating to two distinct episodes of energy release in the flare impulsive phase, and the latter external reconnection led to the deformation of the confined MFR. We propose that the deformation of confined MFRs can complement the understanding of the entire evolution of MFRs and the initiation of solar eruptions. Further observations and simulations are desirable in the future.

{\bf Acknowledgements}

The authors thank the anonymous referee for constructive comments and Dr. Zhenghua Huang for the helpful discussion. We gratefully acknowledge the use of data from Shanghai Astronomy Museum' s Educational Adaptive-optics Solar Telescope (EAST) built by Institute of Optics and Electronics, Chinese Academy of Sciences. The EAST operation is supported by the Shanghai Science \& Technology Museum. {\it SDO} is a mission of NASA's Living With a Star Program. The authors thank the teams of {\it SDO}, SMART, and EAST for providing the data. This work is supported by grants NSFC 11790303, 11727805, and 12073016, and the open topic of the Key Laboratory of Solar Activities of Chinese Academy Sciences (KLSA202108).

\bibliography{sola_bibliography_example.bib}{}
\bibliographystyle{unsrt}

\begin{figure}
\centering
\includegraphics{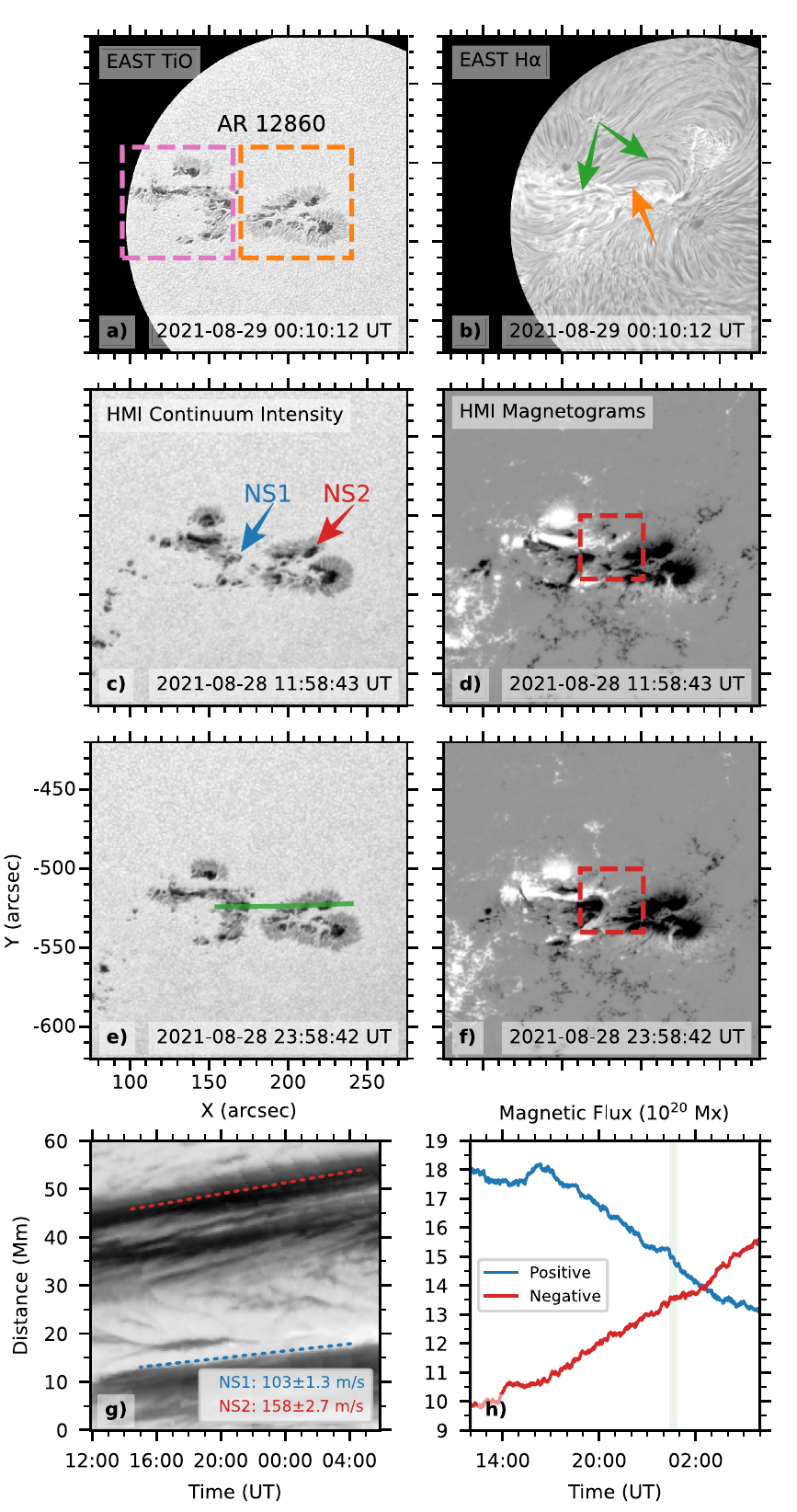}
\caption{The overview of AR 12860 and its magnetic field evolution. ((a)-(f)) The orange and the pink dashed boxes separately outline the preceding sunspots and following sunspots, and the blue and red arrows points out the moving negative sunspots, NS1 and NS2. The green and orange arrows show the sigmoidal filament and the underlying threads. (g) Time-distance plot along the green line in panel (e), showing the movement of NS1 and NS2. The dotted lines are used to derive the attached speeds. (h) The negative (red) and positive (blue) magnetic flux evolution (between 12:00 on August 28 to 06:00 UT on August 29) in the AR center (red boxes in panel (d) and (f)). The green-shaded area represents the duration of the eruption.
\label{f1}}
\end{figure}

\begin{figure}
\centering
\includegraphics{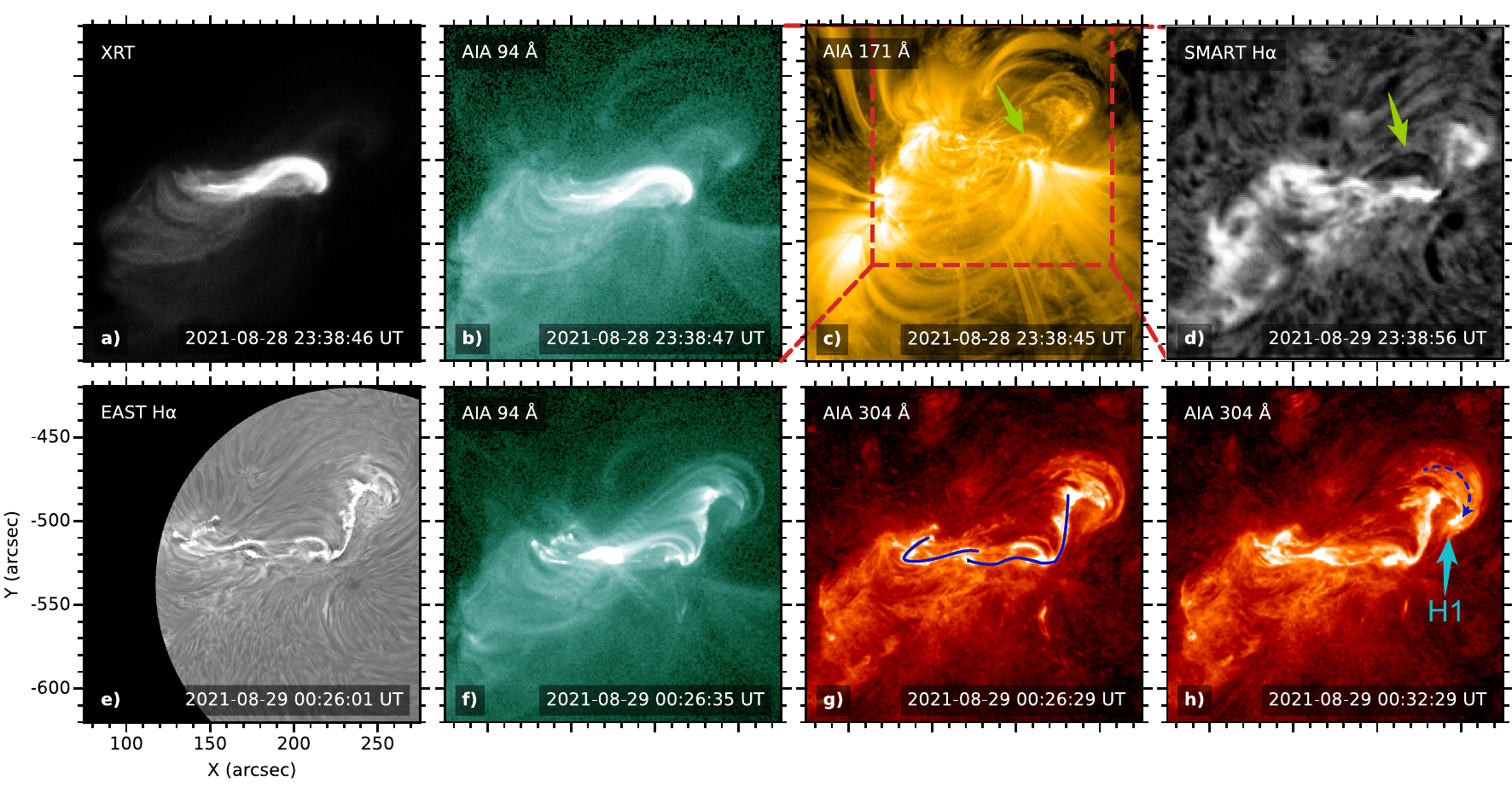}
\caption{The asymmetric MFR before the eruption (upper panels) and flare ribbons after the eruption (bottom panels). The red box points out the field of view of other panels. The curved lines represent the J-shaped and inverse-L-shaped flare ribbons, and the curved arrow shows the clockwise slipping motion. The green and cyan arrows separately indicate the western part of the MFR and the first hook (L1). (An animation is available online, and the animated sequence runs from 23:35 on August 28 to 00:35 UT on August 29.)
\label{f2}}
\end{figure}

\begin{figure}
\centering
\includegraphics{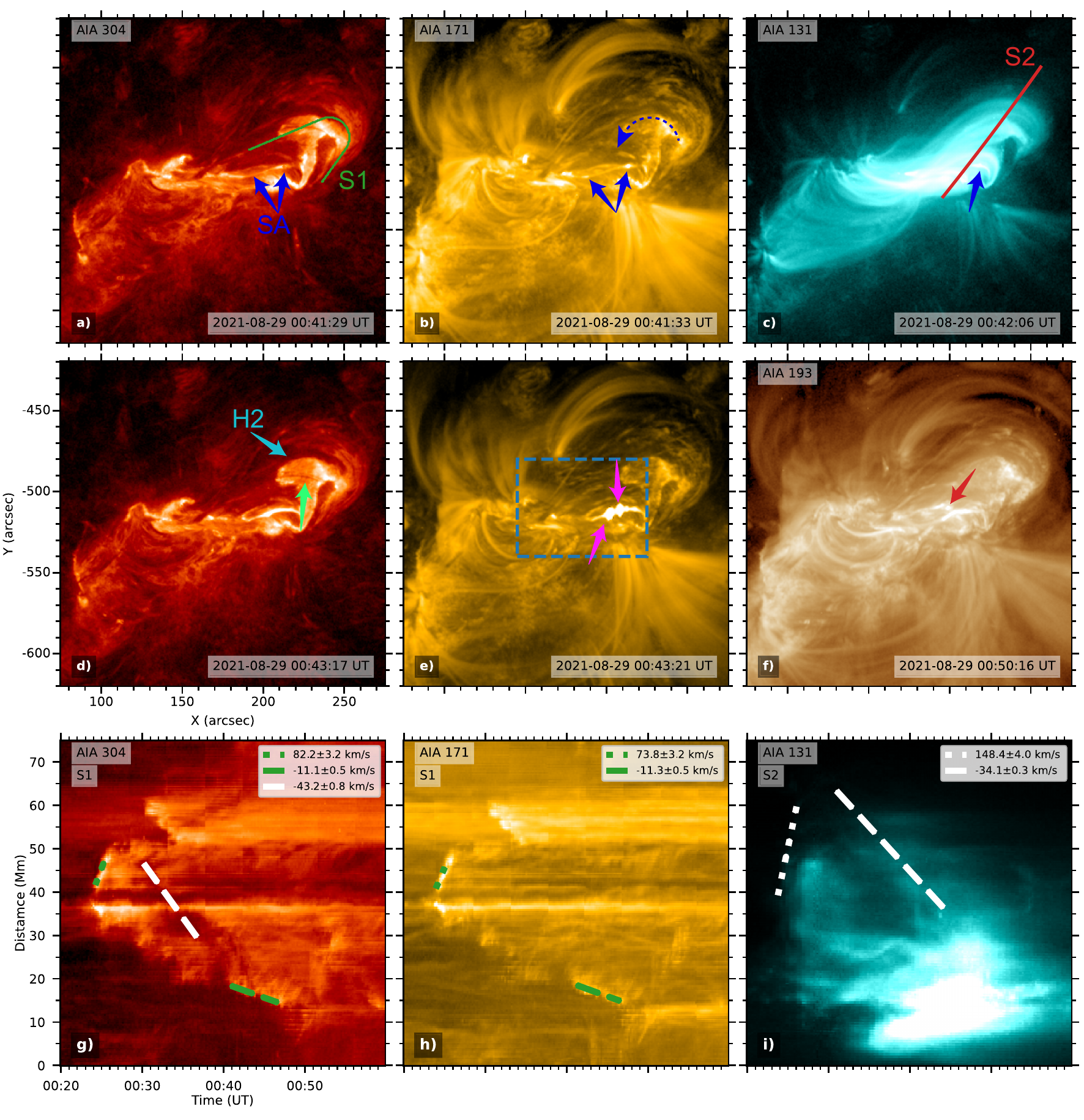}
\caption{((a)-(f)) The deformation of the confined MFR in AIA 304, 171, 193, and 131~{\AA}. The blue arrows indicate the underlying SA, and the pink arrows show the flashings on the SA. The green and cyan arrows point out the shifting filament threads and the second hook (H2), respectively. The red arrow points out the plasmoids. The curved arrow represents the anti-clockwise slipping motion. ((g)-(i)) Time-distance plots along the curve (S1) in panel (a) and the line (S2) in panel (c), uncovering the bi-directional slipping motions and the MFR eruption. The dotted and dashed lines are used to derive the attached speeds. (An animation of AIA images is available online, and the animated sequence runs from 00:35 to 01:00 UT on August 29.)
\label{f3}}
\end{figure}

\begin{figure}
\centering
\includegraphics[width=120mm]{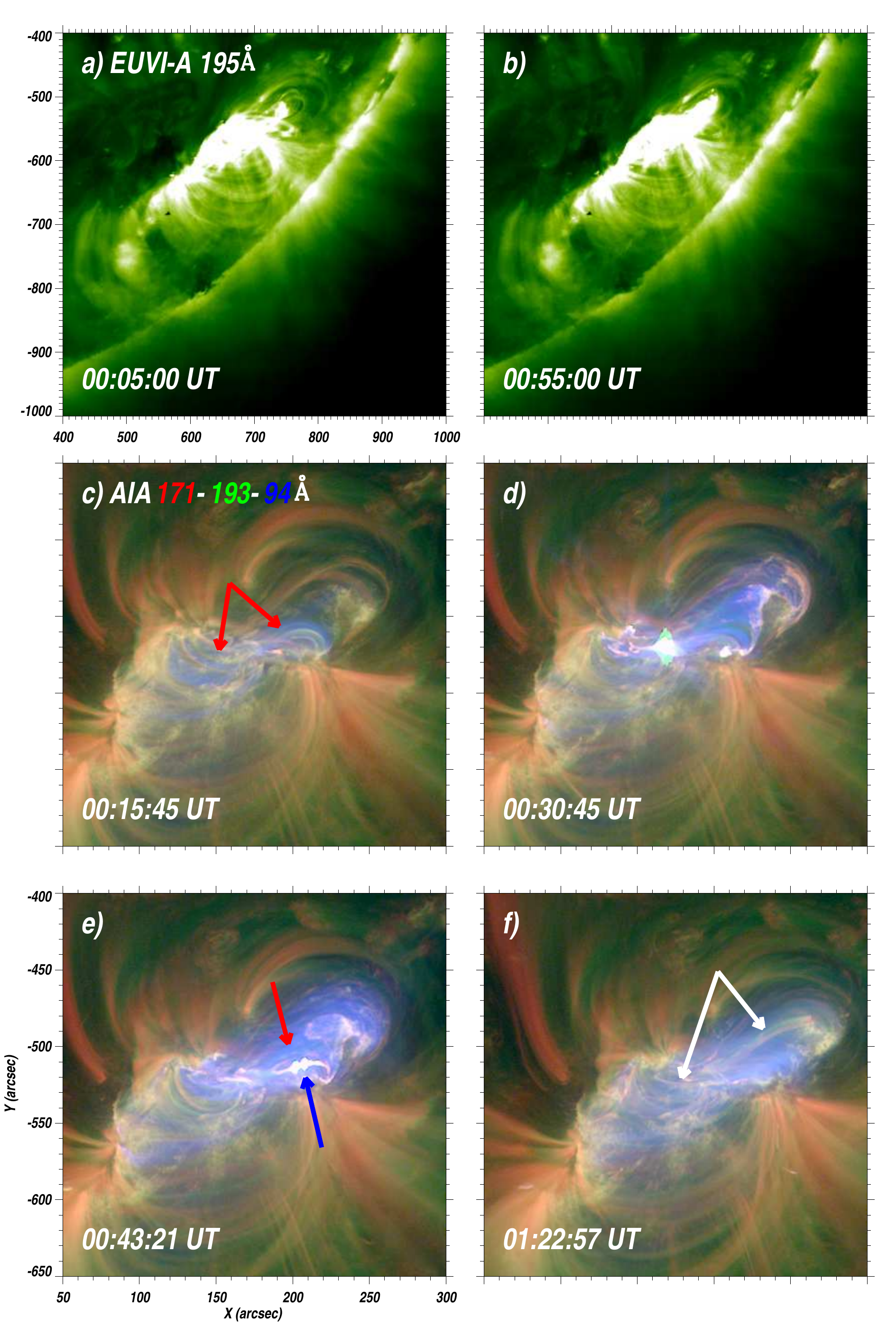}
\caption{The eruption in two perspectives of EUVI-A 195~{\AA} ((a)-(b)) and the composite images of AIA 171 (red), 193 (green), and 94 (blue)~{\AA} ((c)-(f)). The deformation of the confined MFR in AIA 304, 171, 94~{\AA}. The red and blue arrows separately indicate the MFR and the SA. The white arrows show the newly-formed loops after the eruption. (An animation with AIA and EUVI-A images is available online, and the animated sequence runs from 00:00 to 01:00 UT on August 29.)
\label{f4}}
\end{figure}

\begin{figure}
\centering
\includegraphics[width=120 mm]{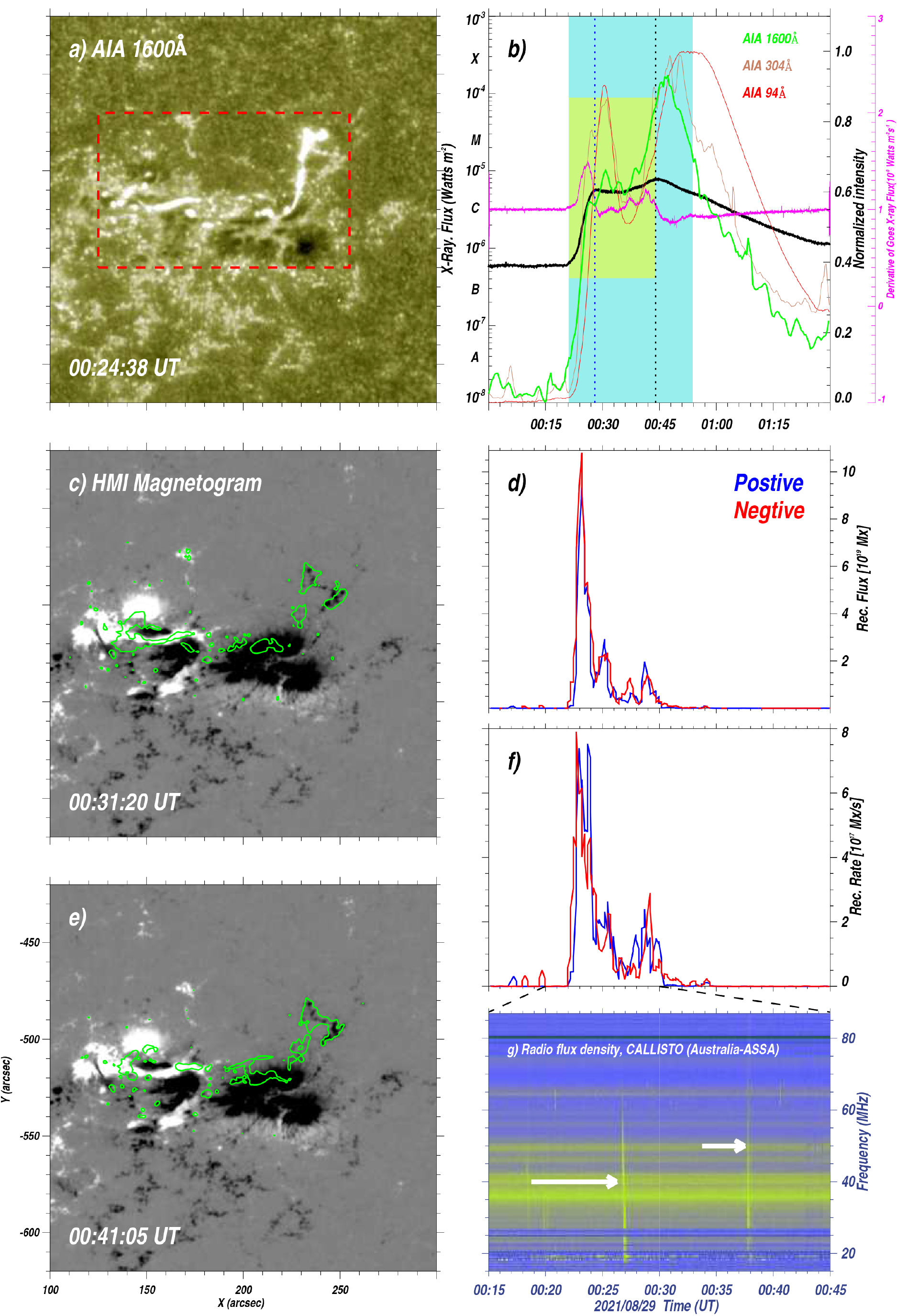}
\caption{The flare ribbons in AIA 1600~{\AA} and their contours over HMI magnetograms (left panels). (b) The intensity curves in AIA 1600, 304, and 94~{\AA} for the eruption region (the red box in panel (a)) and the GOES SXR flux and its derivative for the flare. The vertical lines mark the two peak times of the flare. The yellow and cyan shades separately cover the flare impulsive phase in GOES SXR and AIA 94~{\AA}. ((d) and (f)) The curves of the reconnected flux and the reconnection rate for the positive (blue) and negative (red) magnetic flux. (g) The radio dynamic spectra from CALLISTO, showing the type III radio bursts (white arrows).
\label{f5}}
\end{figure}

\begin{figure}
\centering
\includegraphics[width = 120 mm]{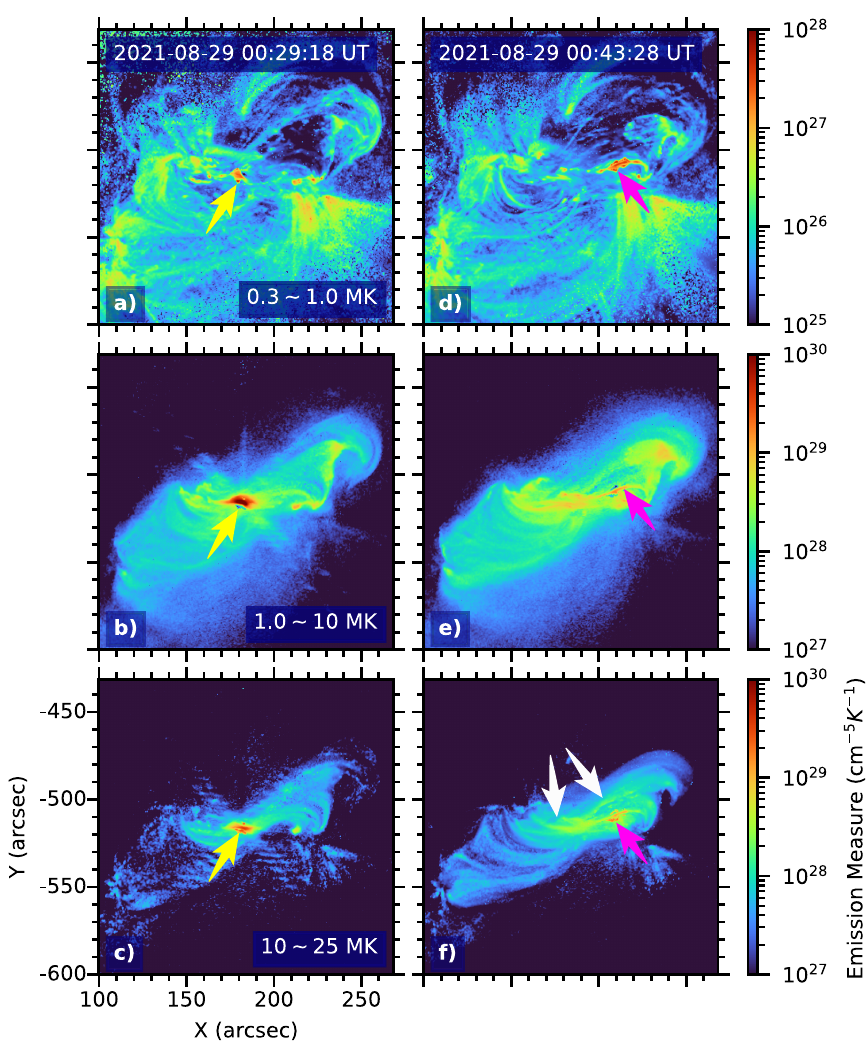}
\caption{Two-step magnetic reconnection in EM maps at temperature range of 0.3-1 MK, 1-10 MK, and 10-25 MK. The yellow and pink arrows separately indicate the post-flare loops and the flashings. The white arrows show the newly-formed high-temperature loops. 
\label{f6}}
\end{figure}
\newpage
\begin{figure}
\centering
\includegraphics{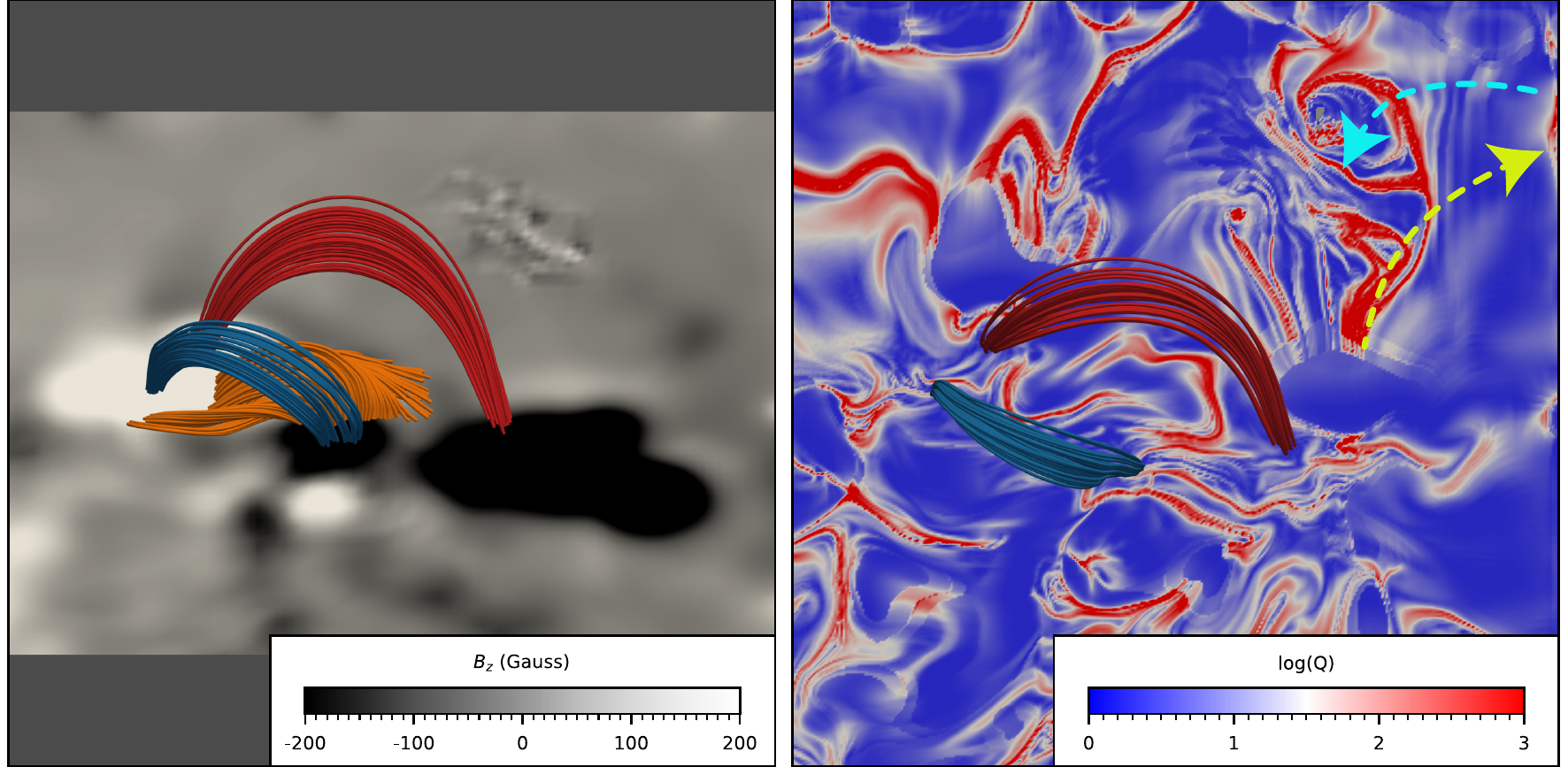}
\caption{The NLFFF extrapolated MFR (orange) and the overlying arcades (blue and red) before the eruption (00:00 UT) superposed on the HMI vector magnetogram  in the side view (left) and on the contours of $Q$ value at photospheric surface in the top view (right). The yellow and cyan dashed arrows show two kinds of slipping motions.
\label{f7}}
\end{figure}
\newpage
\begin{figure}
\centering
\includegraphics[width = 180 mm]{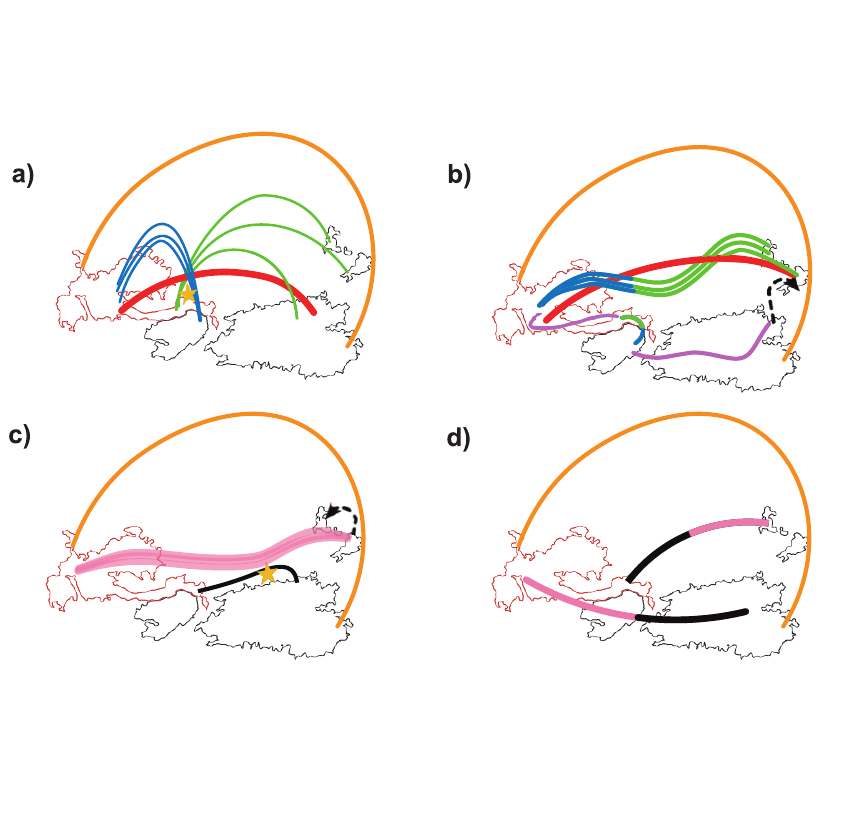}
\caption{Schematic drawings for the eruption and deformation of the MFR (red) below the confining cage (orange) based on the contours of positive and negative polarities from the HMI magnetogram at 00:00 UT on August 29. The green and blue lines represent the filed lines stretched by the rising MFR, and the green-blue lines show the post-flare loops and the newly-formed lines wrapping the rising MFR due to the inner magnetic reconnection. The underlying SA is indicated by the short black line, and the pink-black lines show the newly-formed loops resulted from the external magnetic reconnection. The flare ribbons are outlined by the purple curves, and the two-step magnetic reconnection is indicated by star symbols.
\label{f8}}
\end{figure}

\end{document}